\documentstyle[preprint,prl,aps,tighten]{revtex}
\tighten
\begin{document}
\preprint{}
\draft
%
%
\title{Quantum Zeno Effect\\ 
with the Feynman-Mensky Path-Integral Approach}
\author{Roberto Onofrio}
\address{Dipartimento di Fisica ``G. Galilei'' and INFN, 
Universit\`a di Padova,\\
Via Marzolo 8, Padova, Italy 35131}
\author{Carlo Presilla}
\address{Dipartimento di Fisica and INFN, 
Universit\`a di Roma ``La Sapienza'',\\
Piazzale A. Moro 2, Roma, Italy 00185}
\author{Ubaldo Tambini}
\address{Dipartimento di Fisica and INFN, Universit\`a di Ferrara,\\
Via Paradiso 12, Ferrara, Italy 44100}
\date{Phys. Lett. A (1993) 135-140}
\maketitle
%
%
\begin{abstract}
A model for quantum Zeno effect based upon an effective 
Schr\"odinger equation originated by the path-integral approach 
is developed and applied to a two-level system 
simultaneously stimulated by a resonant perturbation.
It is shown that inhibition of 
stimulated transitions between the two levels appears 
as a consequence of the influence of the meter whenever measurements 
of energy, either continuous or pulsed,
are performed at quantum level of sensitivity.
The generality of this approach allows to qualitatively understand  
the inhibition of spontaneous transitions 
as the decay of unstable particles, originally presented as a paradox 
of quantum measurement theory.
\end{abstract}
%
%
%
%
\narrowtext
Quantum measurement theory has been developed in the thirties to 
understand the problems which arise when quantum mechanical  
formalism is interpreted and confronted with its macroscopic limit. 
Some of the debates originated by the matching between quantum and classical 
worlds assumed the form of paradoxes and have been discussed in terms 
of ideal experiments.  
This is the case of the so-called quantum Zeno effect 
\cite{KHA,MIS1,MIS2,KRAU,SUDB}, 
i.e., the inhibition of the free evolution of a system subjected to 
continuous measurements. 
In the original example of Misra and Sudarshan an unstable particle whose 
trajectory is continuously monitored should never be observed to decay.
Due to recent technological progress in the measurement of physical quantities,
expecially in quantum optics, in the physics of superconducting 
coherent devices such as SQUIDs, and in experimental gravitation, 
some of these ideal experiments can be actually performed \cite{BRAG}. 
Claims of the observation of quantum Zeno effect have been 
reported by Itano et al. \cite{ITA}, which observed freezing of the 
stimulated transition probability in a two-level system subjected 
to frequent measurements of the population of a level. 

The original interpretation of the authors in terms of quantum Zeno effect 
has been debated both with philosophical considerations on the concept 
of measurement and with detailed calculations \cite{BAL}. 
In this Letter we make use of a measurement
model which allows to discuss the general features of 
a measurement operation independently of the
particular measuring apparatus used to perform it.
As a result of this analysis quantum Zeno effect turns out to be 
just an example of the influence of the meter when measurements are performed 
on a system in the quantum regime of sensitivity.

The model is based upon restriction of the Feynman paths to the 
measurement result by introducing a measure functional in the space 
of the paths \cite{ME1,ME2}. 
In this method the effect of the meter on the measured system is taken into
account giving the output of the measurement and the accuracy with which 
it has been performed. 
No explicit degrees of freedom of the meter are introduced, and this makes the 
considerations quite independent of the particular type of measuring apparatus.
The model has been applied to 
understand the accuracy of measurements of position in non-linear 
systems, monitored in a continuous \cite{MEOP1} and impulsive way \cite{MEOP2}. 
In the case of quantum Zeno effect as investigated in \cite{ITA} 
we are interested to measure the energy 
of a system under the simultaneous effect of an external potential
responsible for stimulated transitions.
This makes the Feynman paths in the phase space a very adequate tool.
Let us suppose that a continuous measurement of energy with result 
$E$ is performed for a time $\tau$ (for simplicity the 
result is considered constant) using an instrument with error 
$\Delta E$. The restriction to the paths around the measurement 
result is obtained by introducing a functional weighting the paths, 
for instance with a Gaussian measure,
\begin{equation}
w_{[E]}=\exp{\left\{ -{\langle(H_0-E)^2 \rangle \over \Delta E^2} \right\} }
\label{WEIG}
\end{equation}
where $\langle ... \rangle$ indicates time-average of the 
argument between $0$ and $\tau$. 
The kernel for the propagation of the system under continuous 
monitoring of its energy is modified by the weight functional to
\begin{equation}
K_{[E]}(q^{\prime},\tau;q,0)=
\int d[q] d[p] e^{ 
{i \over {\hbar}}\int_0^{\tau}[p \dot{q} - H_0(q,p)]dt }
w_{[E]}
\label{PROP1}
\end{equation}
which can be rewritten as the usual kernel of a new 
system in which the effect of the measurement has been 
taken into account through an effective Hamiltonian
\begin{equation}
H_{eff}=H_0-i{\hbar\over{\tau \Delta E^2}}(H_0-E)^2~ .
\label{HEFF}
\end{equation}
The non-Hermitian nature of the effective Hamiltonian is due to 
the selective measurement that restricts the possible future 
results \cite{ME3}. 
According to this selection the state of the measured system 
looses its normalization.
During the measurement the evolution of the system, supposed to be 
in a pure state $|\psi(0) \rangle$ at the beginning of the measurement,
is given by the Schr\"odinger equation
\begin{equation}
i \hbar {\partial \over \partial t} |\psi(t) \rangle = 
H_{eff} |\psi(t) \rangle ~.
\label{SCH}
\end{equation}
Let $|n \rangle$ and $E_n$ be the eigenvectors and the eigenvalues 
of the Hamiltonian $H_0$ of the unmeasured system,  
the state $|\psi(t) \rangle$ of the measured system can be expanded in 
the base $\{ |n\rangle \}$
\begin{equation}
|\psi(t) \rangle =\sum_n c_n(t) |n \rangle ~.
\label{PRO}
\end{equation}
By substituting (\ref{PRO}) in (\ref{SCH}) 
an evolution equation for the coefficients $c_n$ is obtained, 
\begin{equation}
{d c_n \over dt} = 
\left [ -i{E_n \over \hbar} -  {(E_n-E)^2 \over \tau\Delta E^2} \right ] \ c_n
\label{CDOT}
\end{equation}
which is solved to get 
\begin{equation}
c_n(t) = \exp \left\{ - i {E_n \over \hbar} t 
- {(E_n-E)^2 \over \tau \Delta E^2} t \right\} ~ c_n(0) ~.
\end{equation}
We note that the coefficients $c_n$ are suppressed by the real
exponential factor with a time-constant $\tau \Delta E^2/ (E_n-E)^2$, 
i.e., the state $|\psi(t) \rangle$ collapses around the measurement result.
This last is not necessarily an eigenvalue $E_n$ of the 
unmeasured system due to the classical uncertainty of the meter \cite{HAAKE}. 
If the measurement result is some definite eigenvalue $E_m$ and
$\Delta E \ll E_n-E_m$ the state of the measured system collapses 
to $|\psi(\tau) \rangle = c_m(\tau) |m \rangle$ at the 
end of the measurement and its squared norm, namely $|c_m(0)|^2$, is the 
probability associated to the initial state to get the measured result $E=E_m$.

The transition between different levels 
is obtained under the action of an appropriate 
external perturbation $V(t)$ which is added 
to the effective Hamiltonian of Eq. (\ref{HEFF}).
The decomposition of the state $|\psi(t) \rangle$
in terms of the same eigenstates $|n \rangle$ of the unmeasured and unperturbed
system can be used again. 
The evolution equation for the coefficient $c_n$ contains also a 
term proportional to the perturbation strength and nondiagonal in the 
index $n$,
\begin{equation}
{d c_n \over dt} = 
\left [ -i{E_n \over \hbar} - {(E_n-E)^2 \over \tau\Delta E^2} \right ] \ c_n
- i \sum_k {V_{nk} \over \hbar} c_k
\label{CDOTSYS}
\end{equation}
where $V_{nk} = \langle n |V(t)| k \rangle$.

A particularly simple picture is obtained for 
a two-level system with energies $E_1$ and $E_2$.
Assuming a perturbation potential $V_{11}=V_{22}=0$ and
$V_{12} = V_{21}^\ast = V_0 e^{i \omega (t-t_0)}$ with $V_0$ real,
the solution of the system (\ref{CDOTSYS}) is
\begin{eqnarray}
c_1(t) = \exp \left\{- i {E_1 \over \hbar} t 
- {(E_1-E)^2 \over \tau \Delta E^2} t +iqt \right\} 
\nonumber \\ 
\times \biggl[ c_1(0) \cos \left( wt \right) +
{qc_1(0) + e^{i \omega t_0} pc_2(0) \over iw }
\sin \left( w t \right) \biggr]
\label{C1}
\end{eqnarray}
\begin{eqnarray}
c_2(t) = \exp \left\{- i {E_2 \over \hbar} t 
- {(E_2-E)^2 \over \tau \Delta E^2} t -iqt \right\} 
\nonumber \\ 
\times \biggl[ c_2(0) \cos \left( wt \right) - 
{qc_2(0) - e^{-i \omega t_0} pc_1(0) \over iw }
\sin \left( w t \right) \biggr]
\label{C2}
\end{eqnarray}
where $p=V_0/\hbar$, $2q=\omega - (E_2-E_1)/\hbar + 2i \Omega$ with
$\Omega = [(E_2-E)^2 - (E_1-E)^2] / 2 \tau \Delta E^2$ 
and $w=\sqrt{q^2+p^2}$.
In order to evidence the Zeno effect in a specific example 
let us suppose initially
the system to be in the state $|1 \rangle$, the perturbation to be
resonant, i.e., $\hbar \omega = E_2-E_1$, and the result of 
the continuous measurement to be $E=E_1$.  
The probability $P_1(t)$ to find the system at time $t \leq \tau$ in the state 
$|1 \rangle$ is 

\begin{equation}
P_1(t) = { |c_1(t)|^2 \over |c_1(t)|^2 + |c_2(t)|^2 }  
= {1 \over 1 +
\left | { V_0/\hbar \over \Omega + w \cot(wt) } \right |^2 }
\label{TRAN}
\end{equation}
where 
\begin{equation}
w=\sqrt{ \left( {V_0 \over \hbar} \right )^2 - \Omega^2 }
\end{equation}
and
\begin{equation}
\Omega = {(E_2-E_1)^2 \over 2 \tau \Delta E^2} .
\end{equation}
A three-dimensional plot of $P_1$ versus 
time and measurement error $\Delta E$ is shown in Fig. 1. 
When the measurement error is large we have $V_0/\hbar \gg \Omega$ 
and the system oscillates between levels 1 and 2 with Rabi frequency 
$2 V_0/\hbar$. 
In the opposite limit of accurate measurements, when $w$ is imaginary,   
an overdamped regime is achieved in which transitions are inhibited.
A critical damping is observed when $w=0$. 
In this case the probability
$P_1$ approaches asymptotically  to the value $1/2$ when $\tau \gg \hbar/V_0$.  
This critical damping is obtained at a measurement error
\begin{equation}
\Delta E_{crit} = (E_2-E_1) \sqrt{ \hbar \over 2 V_0 \tau}
\label{DECR}
\end{equation}
which defines the borderline between the Rabi-like behaviour ($\Delta E >
\Delta E_{crit}$) and the Zeno-like inhibition ($\Delta E<\Delta E_{crit}$) 
in terms of the instrumental accuracy of the meter.
It should be observed that also in the Rabi regime the 
effect of the meter already appears as a deformation of the 
simple harmonic law for the probability $P_1$ with an evident
change in its Fourier spectrum. 

In Ref. \cite{GAG} plots similar to Fig. 1 were obtained 
integrating the optical Bloch equations for a three-level system 
and extracting the evolution of a two-level subsystem. 
In the same paper the effect of the measurement was expressed 
in terms of the coupling to a perturbation term.
Here we have a more general picture of the meter which is taken into account 
by simply giving the measurement result and the accuracy with which 
it has been determined. 

We want also to point out that, although for 
simplicity the case of a constant string of results equal to 
$E_1$ has been considered in our specific example, the formalism 
allows to deal with any possible result $E(t)$ realizing a 
possible history of the energy measurement 
associated to a single quantum 
trajectory of the system \cite{GAG,GRI}. 

The model described here allows also to recover the case of 
measurement pulses which is the subject of the experiment 
described in Ref. \cite{ITA}. 
Let us consider the transition probability 
$P_{1\rightarrow 2}=1-P_1(T)$ at the end of an on-resonance $\pi$ 
pulse of duration $T=\pi V_0/2\hbar$ such that $P_{1 \rightarrow 2}=1$  
without measurements. 
During this time interval $n$ pulsed measurements,
each lasting $\Delta \tau$, can be performed on the system
for a total measurement time $\tau=n \Delta \tau$.
For example we choose a measurement strategy such that the $k$-th
measurement pulse is on from $ kT/n$ to $kT/n + \Delta \tau$
and $k=0,1,...,n-1$. 
Moreover we choose $\Delta \tau/T=10^{-2}$ so that 
the number of measurement pulses which can be performed ranges between 
$1\leq n\leq 100$, for $n=100$ the case of a continuous measurement 
being recovered. 
By iterating $2n$ times Eqns. (\ref{C1}) and (\ref{C2}) alternatively with 
finite $\Delta E$ (during the measurement time interval) and infinite 
$\Delta E$ (during the following non-measurement time interval) we 
obtain the results shown in Fig. 2. 
Independently upon the number of measurement pulses the Rabi-like 
behaviour is observed in the classical regime when
$\Delta E > \Delta E_{crit} \equiv (E_2-E_1)\sqrt{\hbar/2V_0T}$. 
In the quantum regime $\Delta E< \Delta E_{crit}$ the Zeno-inhibition 
increases with the number of measurement pulses until the ultimate 
limit given by the continuous measurement is achieved. 
This also shows that there is no contradiction between continuous 
and discrete measurements because in both the cases inhibition 
of the evolution is achieved provided the measurement is 
sufficiently accurate to perturb the system.

In all these considerations there is no paradox: 
the observed system is 
coupled both to the external perturbation, characterized by the 
Rabi frequency $2V_0/\hbar$, and to the measurement system, 
characterized by the frequency $\Omega$. 
Both the perturbation and the meter 
compete between them to influence the evolution of the observed system. 
Indeed one can choose also quantum measurement strategies 
which make the effect of the meter negligible. 
For instance an impulsive measurement of energy each period of the 
Rabi oscillations should result in measurements of the 
occupation probability at the same instants of time unaffected by the meter 
itself, therefore realizing a Quantum Non-Demolition monitoring 
of the population in level 1 \cite{BRAG,MEOP2,CAVES}.

The crucial role played by the meter accuracy
in a measurement process also applies to 
the case of transitions due to spontaneous emission. 
In the original example of Ref. \cite{MIS1} 
the track of an unstable particle is observed and 
the paradoxical inhibition of its decay is raised. 
Due to the poor spatial resolution in the knowledge of the particle track
it is not surprising that quantum effects of the measurement are negligible, 
i.e., the particle is observed to decay,
despite the microscopic nature of the measured system as 
already pointed out by Landau and Lifshitz \cite{LAN}.
The continuous monitoring of position of a particle schematized as a 
harmonic or an anharmonic oscillator has been discussed \cite{MEOP1}.
It turns out that quantum effects of the measurement are important 
only below a critical accuracy of the order of the De Broglie 
wavelength of the oscillator 
$\Delta x_{crit} \sim \sqrt{\hbar / 2 m \omega}$.
In the case of the decay of elementary particles 
the critical accuracy discriminating the free-decay from the 
Zeno-suppressed decay will be of the order of the Compton wavelength 
of the virtual intermediate vector bosons responsible for the decay.
In the specific example given in Ref. \cite{MIS2} a charged pion decays 
due to the coupling with the electroweak 
vector bosons $W^{\pm}$ which have Compton wavelength 
$\hbar / {M_W c} \simeq 10^{-18}$ m: 
only having a particle detector with spatial resolution of such 
an order of magnitude quantum Zeno effect could be observed, 
appearing as a sort of electroweak microcavity effect.
\acknowledgments 
We are grateful to V. B. Braginsky and G. Ruoso for a critical reading 
of the manuscript.
%
%

%
%
\begin{figure}
\caption{Probability $P_1$ 
to observe a two level system in the state $|1 \rangle$ during the 
measurement interval $0<t<\tau$ and versus the 
normalized measurement error $\Delta E/\Delta E_{crit}$.
The system is in the state $|1\rangle$ at time $t=0$ 
and the result of the measurement between $0$ and $\tau$ 
is $E=E_1$ constant. 
The transition from the underdamped regime with Rabi-like oscillations 
to the overdamped regime with Zeno inhibition is evident when decreasing 
$\Delta E$ below $\Delta E_{crit}$. 
We put $E_2-E_1=V_0=\hbar=1$ and $\tau=2\pi\hbar/V_0$.
\label{FIG1}}
\end{figure}

\begin{figure}
\caption{Transition probability $P_{1 \rightarrow 2}$ at the end of
an on-resonance $\pi$ pulse versus the normalized measurement error
$\Delta E / \Delta E_{crit}$  for pulsed measurements with 
1, 4, 16, 64 pulses and a continuous measurement.}
\label{FIG2}
\end{figure}

\end{document}